\documentclass[twocolumn,showpacs,showkeys,superscriptaddress,prc,nofootinbib]{revtex4-2}
\usepackage{hyperref}
\usepackage[hyphenbreaks]{breakurl}
\hypersetup{breaklinks=true,colorlinks=true, citecolor=blue, urlcolor=blue, linkcolor=blue}
\PassOptionsToPackage{hyphens}{url}
\usepackage{CJK}
\usepackage{mathrsfs}
\usepackage{amssymb}
\usepackage{amsmath}
\usepackage{graphicx}
\usepackage{dcolumn}
\usepackage{bm}
\usepackage{graphicx}
\usepackage{float}
\usepackage[normalem]{ulem}
\usepackage{color}
\usepackage{multirow}
\usepackage{enumerate}
\usepackage{ulem}
\usepackage{textcomp}

\newcolumntype{d}[1]{Dc{.}{.}{#1}}

\begin{document}
	\begin{CJK*}{UTF8}{}
		
		\title{ 
			 The pervasiveness of shape coexistence in nuclear pair condensates
		}
		\author{Y. Lei ({\CJKfamily{gbsn}雷杨})}
		\email[]{leiyang19850228@gmail.com}
		\affiliation{School of Nuclear Science and Technology, Southwest University of Science and Technology, Mianyang 621010, China}
		\author{J. Qi ({\CJKfamily{gbsn}漆洁})}
		\affiliation{School of Mathematics and Physics, Southwest University of Science and Technology, Mianyang 621010, China}
		\author{Y. Lu ({\CJKfamily{gbsn}路毅})}
		\email[]{luyi@qfnu.edu.cn} 
		\affiliation{College of Physics and Engineering, Qufu Normal University, Jingxuan West Road 57, Qufu, Shandong 273165, China}
		\author{H. Jiang ({\CJKfamily{gbsn}姜慧})}
		\affiliation{School of Science, Shanghai Maritime University, Shanghai 201306, China}
		\author{Z. Z. Qin ({\CJKfamily{gbsn}覃珍珍})}
		\affiliation{School of Mathematics and Physics, Southwest University of Science and Technology, Mianyang 621010, China}
		\author{D. Liu ({\CJKfamily{gbsn}刘登})}
		\affiliation{School of Mathematics and Physics, Southwest University of Science and Technology, Mianyang 621010, China}
		\author{Calvin W. Johnson}
		\email[]{cjohnson@sdsu.edu} 
		\affiliation{Department of Physics, San Diego State University, 5500 Campanile Drive, San Diego, California 92182, USA}
		\date{\today}
		
		\begin{abstract}
			We investigate nuclear shape coexistence for a wide range of even-even nuclides. By varying general pair condensates, which include Slater determinants as a limit but also allow for arbitrary pairing channels, we frequently find multiple coexisting mimina, and often more than two. This is consistent with recent experimental results. In order to measure general pairwise correlations beyond a simple Slater determinant, we introduce a novel entropy-like measure, which is smallest mid-shell and largest near shell closures; this is consistent with a picture of pairing-like behavior dominating near closed shells and deformation mid-shell. After surveying nuclides  spanning from the $sd$ shell to nuclides between magic numbers 50 and 82, we focus on the six   lightest nuclei with shape coexistence. Angular-momentum projected variational pair condensate (PVPC) calculations identify band structures, including two newly proposed coexisting bands in $^{26}$Si/Mg and $^{24}$Si/Ne. The PVPC results  agree well with data, providing robust experimental support for the pervasiveness of coexistence in these light nuclei.
		\end{abstract}
		\maketitle
	\end{CJK*}
	\section{Introduction}\label{sec-int}
	The atomic nucleus is a  complex quantum many-body system, characterized by a diversity of structures, such as pairwise correlations, shape deformations, and even shape coexistence. 
	Initially observed six decades ago \cite{PhysRev.101.254.coex.first}, shape coexistence has posed significant challenges to both nuclear experimental techniques \cite{coexistence_rev_exp} and theoretical modeling \cite{coexistence_rev_the_rmp}. 
	
	Recent advancements in radioactive beam  technologies have substantially broadened our understanding of nuclear coexistence \cite{coexistence_rev_exp}. While coexistence was originally observed predominantly near magic numbers and the $N\approx Z$ region (cf.   Fig. 8 of \cite{coexistence_rev_the_rmp}), it is increasingly plausible to suggest that coexistence ``occurs in all (but the lightest) nuclei" \cite{coexistence_rev_the_rmp}, although ``it remains to be seen whether or not this view is correct" \cite{Wood.2016.coex}. {\color{blue} The progress on the experimental evidence for shape coexistence in regions where it has been recently suggested to emerge \cite{coexistence_rev_exp} seems to agree with this view.}
	
	Beyond the pervasiveness of shape coexistence  throughout the nuclear chart, experimental evidences has emerged for multiple (more than 2) coexisting shapes since 2000, when the famous triple coexistence was reported for $^{186}$Pb \cite{coexistence_nature_186Pb}.  Multiple shapes in Ni isotopes were suggested via the comparisons between experimental levels and Monte Carlo shell model calculations \cite{Ni66-mul-co,Ni64-mul-co}. Singh $et~al.$ \cite{Zr98-mul-co} proposed multiple shape coexistence for $^{98}$Zr, although the same level scheme could also be interpreted as phonon excitation~\cite{Zr98-other-no-mul-co}. In $^{110,122}$Cd, up to four distinct shapes are proposed with low-lying $0^+$ states \cite{Cd-mul-co,Cd-mul-co-prc}. 
	
	In recent years, systematic theoretical studies of shape coexistence situated away from magic numbers and $N\approx Z$ regions have been carried out,  employing covariant density functional theory \cite{LZP_coex_PhysRevC.103.054321,greek_coex_PhysRevC.106.044323,ZPW_coex_PhysRevC.107.024308} and machine learning techniques trained with no-core shell models \cite{ML_coex_PhysRevC.105.034306}. Additional island of coexistence were proposed near $N=60,~Z=39$ and $N=90,~Z=66$, based upon cross-parity particle-hole excitations across harmonic oscillator shell gaps at $N=70$ and 112, respectively \cite{greek_coex_PhysRevC.106.044323}. Similarly, M\"oller, \textit{et al}., employed macroscopic-microscopic model to globally investigate the occurrence of shape isomers \cite{global_shape_isomer}. Martinou \textit{et al}. proposed a dual-shell mechanism based on the nature of like-nucleon quadrupole interaction, and suggested several magic numbers for shape coexistence \cite{Martinou2021,Martinou2023}.  Moreover, multiple shape coexistence emerges from different theoretical calculations, including energy density functional calculations \cite{Nomura2017,80zr_5_shapes}, beyond-mean-field calculation with complex excited VAMPIR approach \cite{Petrovici2012}, symmetry-based calculations with partial dynamical symmetry \cite{Leviatan2016,Leviatan2017}, etc.. 
	
	Ideally, an investigation into shape coexistence should use a theoretical framework  accommodating nuclei from light to heavy mass regions, which means it should include pairing correlations as well as shape deformation. Towards this end we use variational pair condensates (VPC)~ \cite{pcv-lei},  a number-conserving scheme which includes Slater determinants, i.e., Hartree-Fock, as a limit~\cite{fu-hfpair}, but also pairwise correlations.
	
	To measure the deviation of the pair condensate from a pure Slater determinant, from the one-body density matrix we compute an entropy-like quantity. This `one-body entropy' follows our expected intuition: it is largest near shell closures, where we expect pairing behavior to dominate, and smallest mid-shell, where one expects shape deformation to dominate. 
	
	To  illustrate the impact of coexistence on low-lying structure, after minimization we project out states of good angular momentum (projected VPC, or PVPC), to generate low-lying level schemes and electric quadrupole transition rates for the six lightest nuclei exhibiting two-shape coexistence from the VPC calculations. Through this process, we identify band structures arising from each shape or their mixing, employing the projection of eigen-states in each subspace with distinct shapes. The observed agreement between our theoretical results and experimental data not only validates the reliability of our calculations but also emphasizes the effectiveness of (P)VPC in capturing the experimental manifestations of coexistence, beyond mere schematic explorations.
	
	This paper is organized as follows: In Sec. \ref{sec:vpc}, we provide an overview of the VPC framework, along with its two variants: TRVPC, which imposes time-reversal constraints, and the Hartree-Fock (HF) limit, which excludes pair correlations. Section \ref{sec:min} presents the distribution of minimum numbers across the nuclear chart up to the 50-82 region, with a detailed analysis on their entropy. In Sec. \ref{sec:band}, we conduct PVPC analysis on the six lightest nuclei exhibiting two shapes, including shape assignments and comparison of low-lying structures between calculations and experiments. Finally, our conclusions are summarized in Sec. \ref{sec:sum}.
	
	\section{Variational pair condensates}\label{sec:vpc}
	
Deformation, pairing, and the variational principle all loom large in nuclear structure theory~\cite{ring_book}.
 The historical success of deformed-variational models incorporating pairing, such as the Bardeen-Cooper-Schrieffer (BCS) with Nilsson basis~\cite{bcs-1,bcs-2,bcs2nucl-2,bcs2nucl-3,bcs2nucl-bohr}, and the Hartree-Fock-Bogoliubov (HFB) model~ \cite{hfb-1,hfb-2}  and its variants, 
inspired other, number-conserving approaches.  The nucleon-pair approximation (NPA) \cite{npa-rev-zhao} uses  collective pairs to truncate the shell model space; the $SD$ pairs in the NPA correspond to the $sd$ bosons in the interaction boson model \cite{ibm}, approximating the quadrupole collective subspace. Another example is  the number-projected BCS model employed to investigate coexistence in $^{66}$Ge, $^{68}$Ge, and $^{68}$Se \cite{hf-yu}.

		Building upon these successes, the variational pair condensate (VPC) model \cite{pcv-lei} is a number-conserving approach which encompasses all possible pairing degrees of freedom to fully demonstrate the pairing effects in nuclear low-lying states. 
 VPC  was applied to $^{132}$Ba \cite{pcv-lei}, successfully identifying the structure of the neutron-excited $10^+$ isomer, which plays a key role in the $10^+$ prolate-oblate coexistence of $^{132}$Ba due to the competition between proton-neutron $(h_{11/2})^2$ alignments \cite{132Ba_h11}.				
		 Adding angular-momentum projection into the VPC framework (PVPC) \cite{pvpc-lu}, employing linear algebra projection \cite{lap1,lap2} after variation, generates low-lying level schemes and transition strengths.  Benchmark calculations of PVPC in the $sd$ shell further confirm its suitability for systematic investigations \cite{pvpc-lu}. Hence, we use the VPC and PVPC to explore the pervasiveness of shape coexistence.
	
	\subsection{Basic formalism}
	
	The VPC adopts a general pair condensate ansatz:
	\begin{equation}\label{eq:vpc:wave}
		\left.\left.\left(\Omega^\dag\right)^N\right| 0 \right\rangle,
	\end{equation}
where $| 0 \rangle$ is a suitable core and 
	 $\Omega^\dag$ is a ``collective pair'' defined as
	\begin{equation}
		\Omega^{\dagger}=\frac{1}{2}\sum_{ij}\omega_{ij} a^{\dagger}_ia^{\dagger}_j.
	\end{equation}
	The creation operators, $a^{\dagger}_i$ and $a^{\dagger}_j$, correspond to the $i$th and $j$th single-particle orbits in the spherical basis, respectively. The $\omega_{ij}$ are structure coefficients of the $\Omega$ collective pair, optimized via the variation 
	\begin{equation}\label{eq:min} 
		\delta\left(\frac{\left\langle \left(\Omega\right)^N \right|\hat H \left| \left(\Omega^{\dagger}\right)^N \right\rangle}{\left\langle  \left(\Omega\right)^N \left|\left(\Omega^{\dagger}\right)^N\right.  \right\rangle}\right)=0, 
	\end{equation} 
	to minimize the energy~\cite{pcv-lei}.
	
	 The particle-number projection of  HFB wave functions can be reformulated similar to the VPC. The particle-number projection of a conventional HFB ansatz \cite{ring_book} can be expressed as
	\begin{equation}
		\begin{aligned}
			\left.\left.\hat P^N\right|{\rm HFB}\right\rangle&=\left.\left.\hat P^N\exp\left(\sum_k\frac{u_k}{v_k}c^\dagger_k c^\dagger_{\overline k}\right)\right| 0 \right\rangle\\
			&=\frac{1}{N!}\left.\left.\left(\sum_k\frac{u_k}{v_k}c^\dagger_k c^\dagger_{\overline k}\right)^N\right| 0  \right\rangle,
		\end{aligned}
	\end{equation}
	where $\hat P^N$ denotes the particle-number projection operator, and $c^\dag_k$ and $c^\dag_{\overline k}$ represents a specific single-particle creation operators and time-reversed partner in the canonical basis, respectively. By transforming from the canonical basis to the spherical basis, $c^\dag_k=\sum_i D_{ki} a^\dag_i$, the projected HFB ansatz becomes
	\begin{equation}\label{eq:hfb}
		\left.\left.\hat P^N\right|{\rm HFB}\right\rangle\propto\left.\left.\left(\frac{1}{2}\sum_{ij}\omega_{ij} a^{\dagger}_ia^{\dagger}_j\right)^N\right| 0 \right\rangle=\left.\left.\left(\Omega^\dag\right)^N\right| 0\right\rangle,
	\end{equation}
	with $\omega_{ij}=2\sum_k \frac{u_k}{v_k}D_{ki}D_{\overline{k}j}$ and $\Omega^\dag=\sum_k\frac{u_k}{v_k}c^\dagger_k c^\dagger_{\overline k}$. This  closely resembles the VPC ansatz of Eq. (\ref{eq:vpc:wave}), except that it requires the $\Omega^\dag$ to be time-reversal invariant, as $\mathcal T\Omega^\dag \mathcal T^{-1}= \sum_k\frac{u_k}{v_k}\mathcal Tc^\dagger_k\mathcal T^{-1} \mathcal T c^\dagger_{\overline k}\mathcal T^{-1} =-\sum_k\frac{u_k}{v_k}c^\dagger_{\overline k} c^\dagger_{ k}=\Omega^\dag$. Here, $\mathcal T$ is the time-reversal operation. Thus, a common HFB variation after particle-number projection corresponds to the VPC with time-reversal invariance. Alternatively, this implies general VPC allows more degree of pairing freedom than conventional HFB.

	\subsection{Constrained VPC}
	
	In addition to exploration with the general VPC, we also investigate the effects of imposing specific constraints on the VPC ansatz. 
	
	The first of the constraints we impose is time-reversal (TR) symmetry on the  VPC ansatz, corresponding to the usual HFB variation after particle-number projection. This constrained variation is denoted as ``TRVPC" and can be realized by enforcing
	\begin{equation}\label{eq:tr}
		\omega_{j_im_i,j_jm_j}=(-1)^{j_i-m_i+j_j-m_j}\omega_{j_i-m_i,j_j-m_j},
	\end{equation}
	so that $\mathcal T\Omega^\dag \mathcal T^{-1}= \Omega^\dag$. Here, the constrained $\Omega^\dag$ corresponds to the pair creation operator in the conventional HFB ansatz, as in Eq. (\ref{eq:hfb}). 
	
	The second constraint forbids pairing, resulting in pure Slater determinants, i.e., antisymmetrized products of single-particle states, which can for even numbers of particles be written in the VPC ansatz, with constant pairing strength under Fermi surface ~\cite{fu-hfpair}:
	\begin{equation}\label{eq:hf}
		|{\rm HF}\rangle= \left(\sum_{k\leq k_F} c^\dagger_k c^\dagger_{\overline k}\right)^N\left | 0 \right\rangle=\left.\left.\left(\Omega^\dag\right)^N\right| 0 \right\rangle,
	\end{equation}
	where $\Omega^\dag=\sum_{k\leq k_F} c^\dag_k c^\dag_{\overline k}$, and $k_F$ corresponds to the index of the canonical basis nearest and below the Fermi surface. Because the Hartree-Fock (HF) approximation is simply finding the Slater determinant that minimizes the energy, consequently HF represents another form of constrained VPC, without pairing correlation. Previous work~\cite{sherpa.bench} showed that HF frequently found shape-coexistence, but also provided evidence of the increasing importance of pairing correlations as one goes up in mass.
	
	\subsection{Slater determinants and pairwise correlations}
	\label{sec:SDpairwise}
	
	One can consider a Slater determinant as an independent particle approximation, with only the Pauli exclusion principle enforced by antisymmetry.	Because our pair condensate ansatz is more general than a Slater determinant, we sought a robust measure of the deviation from 
the independent particle approximation. Here it helps to know that for any single Slater determinant $| \psi_{SD} \rangle$, the one-body density matrix,
$\rho^{(SD)}_{ij} = \langle \psi_{SD} | \hat{a}^\dagger_i \hat{a}_j | \psi_{SD} \rangle,$
is idempotent, that is, $\rho^2 = \rho$, or, equivalently, the eigenvalues $\{ \lambda_\alpha \}$ of $\rho$ are either 0 or 1. Eigenvalues between 0 and 1 signal correlations beyond the independent 
particle approximations.  Therefore, taking our cue from quantum information theory,  from the one-body-density matrix for the (unprojected) pair condensate,
\begin{equation}
\rho_{ij}=\left\langle 0 \left|\left(\Omega\right)^N\left|a^\dag_i a_j\right|\left(\Omega^\dag\right)^N\right| 0 \right\rangle,
\end{equation}
we compute the  `entropy' $S = - \mathrm{tr} \rho \ln \rho $.  In order to compare across nuclides, we add the separately normalized proton and neutron contributions:
\begin{equation}\label{eq:entropy}
			\begin{aligned}
				S_{1b}=&S_\pi+S_\nu,\\
				S_{1b}=&-\frac{\sum_i\lambda^\sigma_i \ln \lambda^\sigma_i}
				{ N_\sigma \ln ( M_\sigma / N_\sigma)}
				{\rm ~~with~~}\sigma=\pi ~{\rm or}~ \nu
			\end{aligned}
		\end{equation} 
		where $\lambda^\sigma_i$ is the $i$-th eigenvalue of the proton or neutron density matrix $\rho^\sigma$, $M_\sigma$ is the number of single proton or neutron orbits, and $N_\sigma$ is valence proton or neutron number. Note that $\mathrm{tr} \rho^\sigma = N_\sigma$.
		
		If the VPC ansatz is reduced to an HF determinant, the eigenvalues of the density matrix are either 0 or 1, so that the entropy $S_{1b}=0$. With pairing correlation, the eigenvalue $\lambda^\sigma_i$ can be between 0 and 1, so that $S_{1b} > 0$.  The maximum entropy,
$S_{1b}=2$, is reached when $\lambda^\sigma_i=N_\sigma/M_\sigma$ (hence our normalization), which would correspond to a seniority-zero wave function. Therefore, our one-body entropy $S_{1b}$ quantitatively measures how much a pair condensate differs from an independent-particle Slater determinant.
		
	We can also measure the the impact of pairwise correlations, by constructing a Slater determinant which approximates the pair condensate.
				As described by Eq. (\ref{eq:hfb}), a particle-number-projected HFB wave-function in the canonical basis can be written in the form of VPC ansatz.  Now we invert this ideal to write the pair in 
 the VPC ansatz in the canonical basis as
		\begin{equation}
			\Omega^\dag=\frac{1}{2}\sum_{ij}\omega_{ij} a^{\dagger}_ia^{\dagger}_j=\sum_k \eta_k c^\dag_k c^\dag_{\overline k},
		\end{equation} 
		with $c^\dag_k=\sum_i D_{ki} a^\dag_k$ defining the canonical basis and $\omega_{ij}=2\sum_k \eta_k D_{ki}D_{\overline{k}j}$, and $\eta_k$ corresponding to the pairing strength of $k\overline k$ channel. 
		By ordering the $|\eta_k|$ and setting the largest $N_\sigma/2$ (where $N_\sigma$ is the number of particles of type $\sigma$) values of $\eta_k = \pm 1$ and the rest to zero, thus defining the Fermi surface, we have now constructed, as a 
		Slater determinant as in Eq. (\ref{eq:hf}), an approximate Hartree-Fock state.  
Finally, we compute
\begin{equation}
\Delta E_\mathrm{pair} = \langle \hat{H} \rangle_\mathrm{approx. \, HF} - \langle \hat{H} \rangle_{\rm VPC}.
\end{equation}
This gives the energy gained by generalizing from Slater determinants to a general pair condensate. 
		
	\subsection{Model spaces and interactions}
	
	Our exploration includes four valence single-particle model spaces separated by magic numbers 8, 20, 40, 50, and 82.  They are $sd$, $pf$, 20-50/50-82, and 50-82 spaces. Specifically, 20-50/50-82 here means proton valence space (holes only in our calculations) comprising the $pf$ orbitals plus the $1g_{9/2}$, while neutron valence space includes the $2s_{1/2}$, $2d_{3/2}$, $2d_{5/2}$, $1g_{7/2}$, and $1h_{11/2}$ orbitals; for our calculations we use valence neutrons for $50 <N \leq 64$, while for $66 \leq N < 82$ we have valence neutron holes.   The 50-82 space is applied to nuclides of $50 < Z \leq 64$, $66 \leq  N < 82$, with valence proton particles and neutron holes.
	
	For the $sd$ and $pf$ space we use  Hamiltonian matrix elements from  widely used shell-model interactions,  USDB interaction \cite{usdb}  and  GXPF1A  \cite{gxpf1a}, respectively. For the remainder of the valences spaces we use  a phenomenological shell model interaction described in Appendix B of Ref \cite{npa-cal-zhao}. Since this phenomenological shell model interaction is not finely tuned for these regions, our VPC calculations only schematically demonstrate the possibility of coexistence in middle-heavy or heavy nuclei, in order to access a large and global sample analysis on minima. Since the calculations in the $sd$ space are mostly trustworthy, relying upon thoroughly verified USDB interaction, we conduct more detailed analyses on six of the lightest $sd$ nuclei exhibiting two shapes, in Sec.~\ref{sec:band}.
	
	\subsection{Identifying minima}
	
	In our search for minima we conduct 1000 VPC runs with randomly initialized $\omega_{ij}$ for each nucleus under investigation. To exclude saddle points, for each resultant minimum  we add a small perturbation and repeat gradient descent, to confirm convergence to the same (stable) minimum. 
	
	To be sure we have identified 
 distinct minima rather than, due to numerical error, states near the same minimum, 
 we use three criteria: 
		\begin{itemize}
			\item 
			The energy difference, denoted by $\Delta E_\mathrm{min}$: the difference between $\langle \hat{H} \rangle_\mathrm{VPC}$ of two candidate minima. (Note this is distinct from $\Delta E_\mathrm{pair}$.)
			\item
			The shape difference, defined by
			\begin{equation}\label{eq:ds}
				\begin{aligned}
					\Delta {\rm Shape}=&\left\{\left(\beta_1\cos\gamma_1-\beta_2\cos\gamma_2\right)^2\right.\\
					&+\left.\left(\beta_1\sin\gamma_1-\beta_2\sin\gamma_2\right)^2\right\}^{\frac{1}{2}},
				\end{aligned}
			\end{equation} 
			where $(\beta_1,\gamma_1)$ and $(\beta_2,\gamma_2)$ are the quadrupole deformation parameters calculated following the formalism outlined in Sec IIB of Ref. \cite{deform}. The same formalism is also used in other calculations of quadrupole deformation parameters throughout the paper. $\Delta {\rm Shape}$ corresponds to the distance between minima on the potential energy surface.
			\item 
			The linear dependence between two minima, computed through the overlap between the $0^+$ states projected from the candidate minima.
		\end{itemize}
		Our specific criteria are to have $\Delta E>0.5$ MeV or $\Delta {\rm Shape}>0.05$, with $0^+$ overlap less than 0.9. With these criteria we are confident in our identification of physically distinct VPC minima.


	\section{The landscape of coexistence }\label{sec:min}
	\begin{figure}
		\includegraphics[angle=0,width=0.45\textwidth]{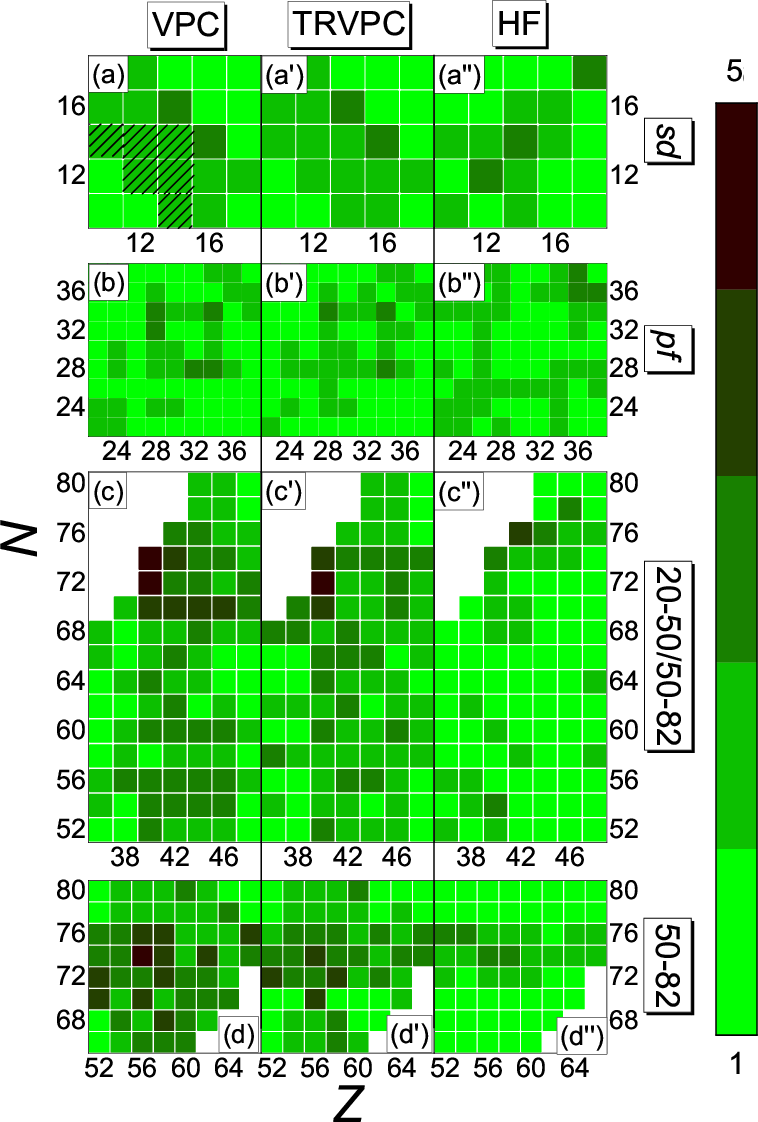}
		\caption{(Color online)
			Number of coexisting minima in different valence spaces, with the legend bar on the right. 
			The three columns of panels correspond to VPC, TRVPC, and HF results, respectively. Experimentally unobserved nuclei are excluded represented with blank space. Each row of panels represents distinct nuclear regions: $sd$, $pf$, 20-50/50-82, and 50-82 major shell. Panel (a) highlights the lightest six nuclei with two minima, as in the cross-hatched area:  $^{24}$Ne, $^{24,26}$Mg, and $^{24\sim 28}$Si, whose wave functions, spectra, and E2 transition rates are  analyzed in Sec. \ref{sec:band}.
		}\label{fig:land}
	\end{figure}

	Figure \ref{fig:land} illustrates the number of minima for nuclides throughout the $sd$, $pf$, 20-50/50-82, and 50-82 valence spaces. Coexistence corresponds to more than one minima.  
In our VPC calculations, coexistence is not rare but rather pervasive. In Table \ref{tab:4min} we list  `super-coexistence' nuclides---those with four or five VPC minima---along with the details of these minima. We remind the reader that experimental evidence for up-to-four shape coexistence states has been reported for $^{110,112}$Cd \cite{Cd-mul-co,Cd-mul-co-prc}.

	\begin{table*}
		\caption{ Energy ($E$ in MeV), quadrupole deformation parameters ($\beta$ and $\gamma$ in deg), and one-body entropy ($S_{1b}$) for each minimum for nuclei with 4 or 5 minima. The minima listed here are obtained with an untuned phenomenological shell model interaction, so there may be no exact experimental correspondences to some of these minima.}\label{tab:4min}
		\begin{tabular}{cccm{2em}<\centering c|cccm{2em}<\centering c|cccm{2em}<\centering c|cccm{2em}<\centering ccc}
			\hline\hline
			 & $E$ & $\beta$ & $\gamma$ & $S_{1b}$ &  & $E$ & $\beta$ & $\gamma$ & $S_{1b}$ &  & $E$ & $\beta$ & $\gamma$ & $S_{1b}$ &  & $E$ & $\beta$ & $\gamma$ & $S_{1b}$ \\
			 \hline
			\multirow{4}{*}{$^{106}$Zr}  & $-$7.3 & 0.108 & 60 & 0.112 &  \multirow{4}{*}{$^{116}$Mo}  & $-$11.7 & 0.000 & 33 & 1.299 &  \multirow{4}{*}{$^{126}$Ba}  & $-$8.9 & 0.187 & 56 & 0.560 &  \multirow{4}{*}{$^{132}$Ce}  & $-$9.2 & 0.033 & 5 & 1.058 \\
			& $-$7.3 & 0.109 & 60 & 0.063 &    & $-$10.9 & 0.014 & 52 & 1.183 &    & $-$8.9 & 0.160 & 22 & 0.671 &    & $-$8.6 & 0.021 & 25 & 0.842 \\
			& $-$7.0 & 0.044 & 58 & 0.120 &    & $-$9.5 & 0.037 & 3 & 1.078 &    & $-$8.0 & 0.173 & 36 & 0.423 &    & $-$7.8 & 0.102 & 8 & 0.645 \\
			& $-$6.5 & 0.024 & 23 & 0.135 &    & $-$8.3 & 0.044 & 45 & 0.890 &    & $-$8.0 & 0.160 & 12 & 0.397 &    & $-$7.8 & 0.057 & 34 & 0.741 \\
			&    &    &    &  &    &    &    &    &  &    &    &    &    &  &    &    &    &    &  \\
			\multirow{4}{*}{$^{110}$Zr}  & $-$8.0 & 0.062 & 2 & 0.521 &  \multirow{4}{*}{$^{114}$Ru}  & $-$12.1 & 0.001 & 48 & 1.318 &  \multirow{5}{*}{$^{130}$Ba}  & $-$10.0 & 0.195 & 0 & 0.939 &  \multirow{4}{*}{$^{134}$Ce}  & $-$8.3 & 0.045 & 37 & 1.107 \\
			& $-$7.9 & 0.155 & 59 & 0.137 &    & $-$11.2 & 0.038 & 28 & 1.161 &    & $-$9.2 & 0.180 & 6 & 0.849 &    & $-$8.2 & 0.228 & 1 & 0.548 \\
			& $-$7.2 & 0.068 & 35 & 0.372 &    & $-$10.0 & 0.026 & 40 & 1.093 &    & $-$8.7 & 0.230 & 1 & 0.523 &    & $-$7.8 & 0.015 & 59 & 0.870 \\
			& $-$5.5 & 0.128 & 52 & 0.285 &    & $-$9.5 & 0.080 & 54 & 0.711 &    & $-$8.1 & 0.153 & 3 & 0.793 &    & $-$7.5 & 0.066 & 38 & 0.946 \\
			&    &    &    &  &    &    &    &    &  &    & $-$7.8 & 0.212 & 2 & 0.196 &    &    &    &    &  \\
			&    &    &    &  &    &    &    &    &  &    &    &    &    &  &    &    &    &    &  \\
			\multirow{5}{*}{$^{112}$Zr}  & $-$9.5 & 0.040 & 16 & 0.924 &  \multirow{4}{*}{$^{116}$Pd}  & $-$8.7 & 0.004 & 52 & 1.323 &  \multirow{4}{*}{$^{132}$Ba}  & $-$8.9 & 0.192 & 0 & 0.989 &  \multirow{4}{*}{$^{136}$Sm}  & $-$7.6 & 0.148 & 2 & 0.454 \\
			& $-$9.3 & 0.118 & 60 & 0.615 &    & $-$8.0 & 0.048 & 9 & 1.234 &    & $-$8.2 & 0.183 & 4 & 0.864 &    & $-$7.3 & 0.158 & 59 & 0.629 \\
			& $-$8.8 & 0.080 & 15 & 0.781 &    & $-$6.9 & 0.116 & 34 & 0.953 &    & $-$7.5 & 0.208 & 4 & 0.527 &    & $-$6.2 & 0.142 & 8 & 0.309 \\
			& $-$8.7 & 0.134 & 55 & 0.579 &    & $-$6.3 & 0.141 & 51 & 0.745 &    & $-$6.9 & 0.143 & 7 & 0.776 &    & $-$5.4 & 0.077 & 43 & 0.616 \\
			& $-$7.3 & 0.083 & 22 & 0.425 &    &    &    &    &  &    &    &    &    &  &    &    &    &    &  \\
			&    &    &    &  &    &    &    &    &  &    &    &    &    &  &    &    &    &    &  \\
			\multirow{5}{*}{$^{114}$Zr}  & $-$9.9 & 0.009 & 4 & 1.055 &  \multirow{4}{*}{$^{122}$Te}  & $-$7.0 & 0.120 & 30 & 1.038 &  \multirow{4}{*}{$^{126}$Ce}  & $-$9.4 & 0.164 & 59 & 0.529 &  \multirow{4}{*}{$^{142}$Dy}  & $-$2.5 & 0.185 & 60 & 0.433 \\
			& $-$9.2 & 0.002 & 45 & 0.854 &    & $-$6.0 & 0.159 & 39 & 0.686 &    & $-$8.5 & 0.117 & 52 & 0.605 &    & $-$2.4 & 0.112 & 19 & 0.730 \\
			& $-$8.7 & 0.126 & 59 & 0.643 &    & $-$5.9 & 0.185 & 58 & 0.458 &    & $-$7.8 & 0.140 & 59 & 0.263 &    & $-$1.8 & 0.105 & 2 & 0.506 \\
			& $-$8.3 & 0.034 & 52 & 0.760 &    & $-$5.7 & 0.144 & 17 & 0.682 &    & $-$7.5 & 0.086 & 10 & 0.578 &    & $-$1.2 & 0.114 & 5 & 0.305 \\
			& $-$7.2 & 0.145 & 25 & 0.381 &    &    &    &    &  &    &    &    &    &  &    &    &    &    &  \\
			&    &    &    &  &    &    &    &    &  &    &    &    &    &  &    &    &    &    &  \\
			\multirow{4}{*}{$^{112}$Mo}  & $-$11.8 & 0.010 & 32 & 1.121 &  \multirow{4}{*}{$^{124}$Te}  & $-$7.5 & 0.105 & 7 & 1.241 &  \multirow{4}{*}{$^{130}$Ce}  & $-$9.6 & 0.029 & 11 & 1.025 &    &    &    &    &  \\
			& $-$10.2 & 0.033 & 23 & 0.924 &    & $-$7.0 & 0.156 & 3 & 1.034 &    & $-$9.1 & 0.200 & 1 & 0.358 &    &    &    &    &  \\
			& $-$9.5 & 0.024 & 28 & 0.840 &    & $-$6.2 & 0.184 & 7 & 0.791 &    & $-$9.0 & 0.111 & 12 & 0.863 &    &    &    &    &  \\
			& $-$8.9 & 0.079 & 55 & 0.563 &    & $-$5.8 & 0.182 & 43 & 0.732 &    & $-$8.3 & 0.081 & 1 & 0.652 &    &    &    &    &  \\
			\hline\hline		
		\end{tabular}
	\end{table*}
	
	Not surprisingly, breaking of symmetries allows for more minima.  Figure \ref{fig:land} shows that VPC can yield more minima than time-reversal-symmetric VPC (TRVPC) and Hartree-Fock (HF) calculations.
	This observation is particularly pronounced around nuclei with five minima, i.e., the $^{112,114}$Zr in the 20-50/50-82 space, and $^{130}$Ba in the 50-82 region.
	A similar result was found in a beyond mean-field calculation of $^{80}$Zr \cite{80zr_5_shapes}, where allowing unrestricted triaxiality was essential for coexistence with up to five minima. 
In our calculations, which also allow for unrestricted triaxiality, inclusion of all pairing channels in VPC enhances coexistence.
	
	We see in Fig. \ref{fig:land} a positive correlation between the inclusion of many pairing channels in our calculations, and the appearance of multiple coexisting states. In general more minima appear in the full VPC calculation, which	allows all pairing channels, than in either the TRVPC calculation,
	which is constrained to time-reversal-invariant
	pairing, or the HF calculations, which have no pairing
	correlations; and TRVPC leads to more minima than
	HF. These trends suggest that, in the same way that
	less-constrained calculations lead to lower minima (e.g., allowing triaxiality leads to lower energies than with constraints to axial symmetry), they also allow for more coexisting minima.
	
Figure \ref{fig:land} also empirically suggests a correlation between the frequency of minima and the number of valence orbitals. Whether this trend holds for superheavy nuclei, where
 the density of single-particle levels increases, remains to be seen.
	
\begin{figure}
	\includegraphics[angle=0,width=0.45\textwidth]{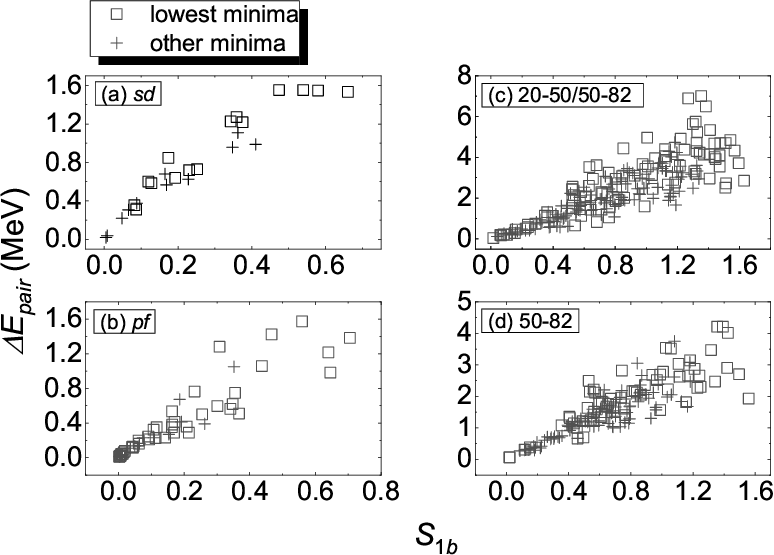}
	\caption{
		$\Delta E_\mathrm{pair}$ against $S_{1b}$. $\Delta E_\mathrm{pair}$ is the energy difference between VPC minimum and corresponding approximate HF wave-function constructed with the canonical bases from the VPC minimum, as discussed in Sec.~\ref{sec:SDpairwise}. $S_{1b}$ is the one-body entropy of VPC minimum defined in Eq. (\ref{eq:entropy}). The hollow squares ($\square$) corresponds to the lowest minima, or VPC ground state, and the crosses ($+$) correspond to other minima.
	}\label{fig:s}
\end{figure}

		Figure \ref{fig:s} displays the correlation between $\Delta E_\mathrm{pair}$, and the one-body entropy $S_{1b}$.  The approximate linear relation is unsurprising, as $S_{1b} \rightarrow 0$ means the 
pair condensate is well approximated by a single Slater determinant (Hartree-Fock-like state). Conversely, we find a VPC minimum with strong pairing generally gains significant energy over the HF-like state. We also find that the largest $S_{1b}$ values in each region always correspond to the lowest minima, marked by $\square$. For other minima, marked by $+$, the $S_{1b}$ values seems systematically smaller, especially in the $sd$ space. Therefore, the VPC ground minimum is likely to exhibit more pairing correlation than higher ones. 
%
%

		\begin{figure}
			\includegraphics[angle=0,width=0.45\textwidth]{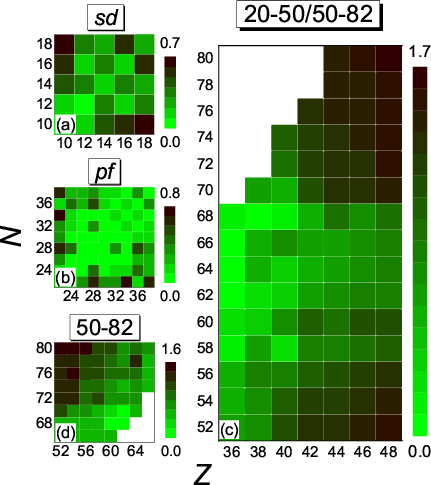}
			\caption{(Color online)
				Distribution of one-body entropy ($S_{1b}$), as defined in Eq. (\ref{eq:entropy}) for the lowest VPC minima.  The legend bar is on the right.
				 Panel (c) is zoomed in to emphasize to the relatively large $S_{1b}$ along with the shell closure at $Z=50$ and $N=50,82$.
			}\label{fig:s_land}
		\end{figure}
		
		We also present the distribution of $S_{1b}$ for the lowest VPC minima in Fig. \ref{fig:s_land}.  It tends to be largest near shell closures, where one expects pairing correlations to dominate over deformation. Conversely, midshell nuclei have strong deformation driving the condensate to a Slater-determinant-like state. 
We also see that, relatively speaking, the entropy, and thus the pairing correlations, are less pronounced in the $sd$ and $pf$ valence spaces than for the 20-50/50-82 and 50-82 spaces, which agrees 
with our expectation that pairing is more important in heavy nuclei than in light nuclei.

	
	\section{coexisting bands}\label{sec:band}
	
	To relate the  minima in Fig. \ref{fig:land} to experimental band structure, we further analyze the six lightest nuclei exhibiting coexistence in VPC calculations. Highlighted in Fig. \ref{fig:land}(a) by diagonal hatching,  these nuclei are $^{24}$Ne, $^{24,26}$Mg, and $^{24\sim 28}$Si. As these are computed using the isospin-conserving USDB interaction~\cite{usdb},  however,  mirror nuclei such as $^{26}$Si/Mg and $^{24}$Si/Ne yield identical results.  Experimental evidence of coexistence in $^{28}$Si has been reported and was theoretically supported \cite{28Si_exp,28Si_phf}. Additionally, $^{24}$Mg has also been proposed as a candidate for a coexistence nucleus \cite{coexistence_rev_the_rmp}.
	In this section we first discuss the VPC minima, and then project out states of good angular momentum to produce excitation bands.
	
	\begin{table}
		\caption{Energy ($E$ in MeV), quadrupole deformation parameters ($\beta$ and $\gamma$ in degrees), and one-body entropy ($S_{1b}$) for each minimum of $^{28}$Si, $^{26}$Si/Mg, $^{24}$Si/Ne, and $^{24}$Mg, highlighted in Fig. \ref{fig:land}(a) with diagonal hatching.}\label{tab:min}
		\begin{tabular}{cccm{1em}<\centering c|cccm{1em}<\centering cccccccccccccccccccccccc}
			\hline\hline									
			&	$E$	&	$\beta$	&	$\gamma$ &$S_{1b}$&&	$E$	&	$\beta$	&	$\gamma$&$S_{1b}$\\
			\hline									
			\multirow{2}{*}{$^{28}$Si}		&	$-$130.1	&0.195		&	60 &0.008	&\multirow{2}{*}{$^{26}$Si/Mg}		&	$-$100.1	&	0.192	&	35&0.169\\
			&	$-$122.8	&0.229		&	0&0.005& &	$-$98.2	&	0.224	&	0&0.049\\
			\hline																		
			\multirow{2}{*}{$^{24}$Mg}		&	$-$81.1	&	0.273	&	10&0.082 &\multirow{2}{*}{$^{24}$Si/Ne}	&	$-$67.7	&	0.162	&	32&0.474	\\
			&	$-$75.7	&	0.211	&	60&0.348&		&	$-$67.1	&	0.171	&	60&0.262\\
			\hline\hline		
		\end{tabular}
	\end{table}
	
	We determine the quadrupole deformation parameters ($\beta$, $\gamma$) for each linearly independent minimum in $^{28}$Si, $^{26}$Si/Mg, $^{24}$Si/Ne, and $^{24}$Mg, and summarize the results in Table \ref{tab:min}, along with their corresponding energies. For $^{28}$Si, we identify two linearly independent minima exhibiting prolate and oblate deformations, respectively, suggesting a potential prolate-oblate coexistence with an energy difference of approximately 7 MeV. This observation is consistent with previous HF calculations, which also detected a prolate band head at 6.69 MeV above the oblate ground \cite{28Si_phf}. 
	
	In the case of $^{26}$Si/Mg, we found an oblate local minimum in addition to a triaxially deformed global minimum at $\gamma\sim 35^\circ$, with an energy difference of  $\sim 2$ MeV. For $^{24}$Si/Ne, we discern two minima characterized as oblate and triaxially deformed, respectively, with a small energy difference of only $\sim$0.5 MeV, likely leading to   mixing.  In $^{24}$Mg, we found  a near-prolate minimum and an oblate minimum, separated in energy by  $\sim$ 5 MeV, maybe leading to a more complex triaxial level scheme compared to $^{28}$Si with its pure prolate-oblate coexistence.
	
\begin{figure}
		\includegraphics[angle=0,width=0.45\textwidth]{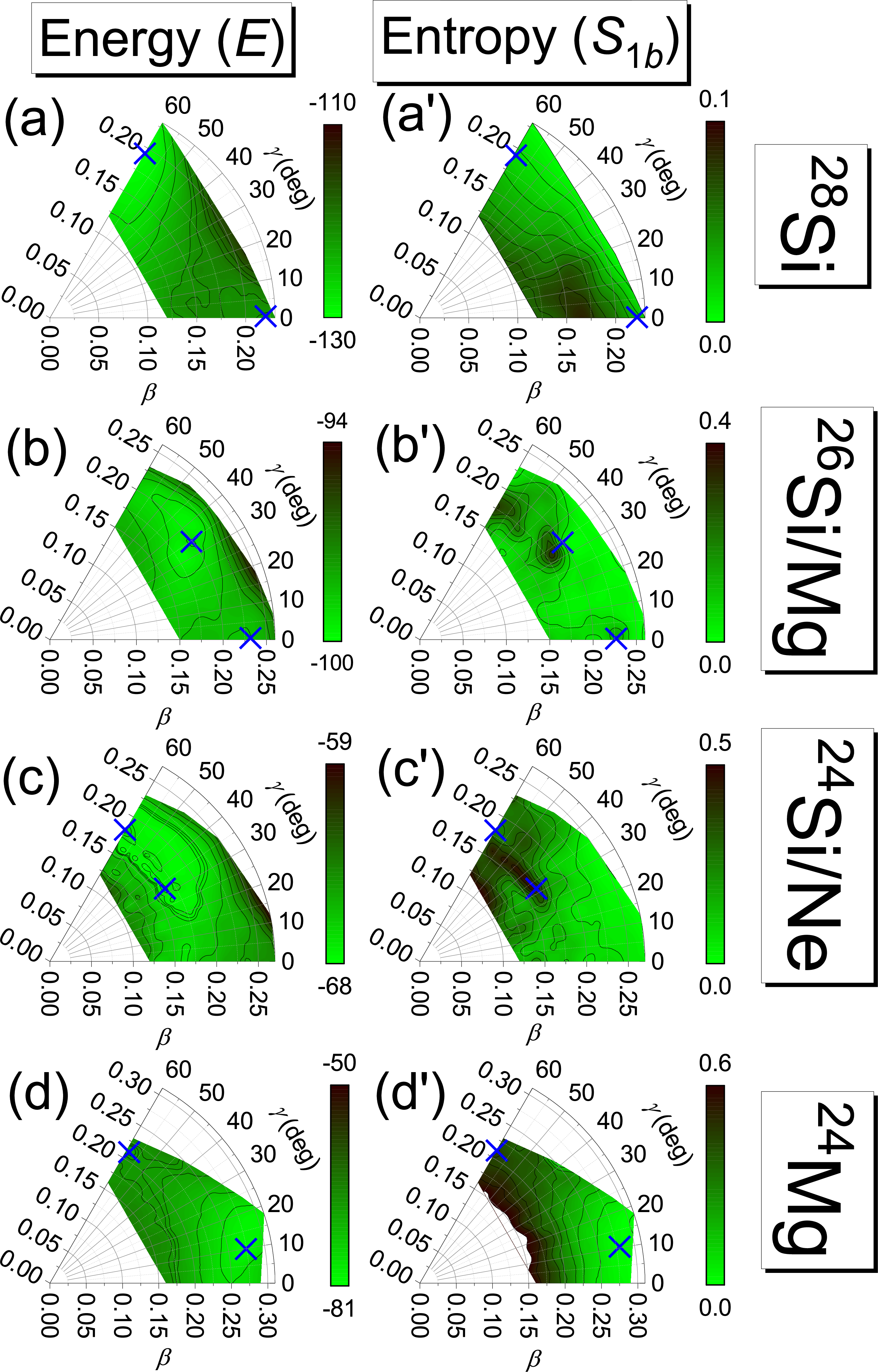}
		\caption{(Color online)
			Energy ($E$) and one-body entropy ($S_{1b}$) surface on the $(\beta,\gamma)$ plane, from shape-constraint VPC, for $^{28}$Si, $^{26}$Si/Mg, $^{24}$Si/Ne, and $^{24}$Mg. The blue crosses ({\color{blue}$\times$}) correspond to the minima listed in Table \ref{tab:min}. $S_{1b}$ is calculated with Eq. (\ref{eq:entropy})
		}\label{fig:pes}
	\end{figure}

	We perform shape-constrained VPC around the minima listed in Table \ref{tab:min}, and present the energy and one-body entropy surfaces in Fig. \ref{fig:pes}. 
For  $^{24}$Si/Ne, we found many minima with $E\sim -67.7$ MeV, $\beta\sim 0.15$ and $\gamma$ in the range of $30\sim 60^\circ$ [Fig.~\ref{fig:pes}(c)]; further investigation, however, 
found strong linear dependencies for the angular-momentum-projected $0^+$ states, so we only take the $\gamma\sim 30^\circ$ minimum as  representative in Table \ref{tab:min}. This suggests gamma softness. Such gamma-soft minima also correspond to larger one-body entropies as shown in Fig. \ref{fig:pes}(c'). 
		
	Generally,  we found no direct correlation between the energy surfaces and the corresponding entropy surfaces in Fig. \ref{fig:pes}. For $^{28}$Si, both minima are found at low entropy; this is unsurprising, given that $^{28}$Si is in the middle of a major shell, where pairing correlations are weaker, as seen in Fig. \ref{fig:s_land}. For $^{26}$Si/Mg,  both minima are located near  peaks in the entropy surface, yet nonetheless have relatively small entropies. For $^{24}$Si/Ne, the minima are near a ridge of near-constant, moderate entropy correlated with the previously noted gamma softness. Finally, for $^{24}$Mg, the entropy surface is  similar to the energy surface; the lowest minima is HF-like, i.e., nearly zero entropy, while the other minimum has a larger entropy. Nevertheless, 
$\sim$50\% of minima are associated with larger entropy;  thus proper treatment of pairing is needed to thoroughly probe nuclear coexistence, which we also found in Fig. \ref{fig:land}.
	
	So far we have only focused on the minima. In order to investigate excitation bands, we must project out states of good angular momenta (PVPC), specifically
	from the minima listed in Table \ref{tab:min}.  We label each distinct pair-condensate minimum by $\omega$, which can be thought of as a proxy for the deformation parameters $(\beta, \gamma)$ listed in Table \ref{tab:min}, as well as 
any other quantity such as the single-particle entropy.  One can expand such a state
\begin{equation}
| \omega \rangle = \sum_{J,K} \psi_{J,K} | \omega, J K \rangle,
\end{equation}
where $J$ is the angular momentum, and $K$ is the angular momentum projection onto the body-frame axis. From this we project out states of good angular momentum $J$ and $z$-component $M$,
\begin{equation}
| \omega, J (K) M \rangle = \hat{P}^J_{M,K} | \omega \rangle, \label{projection}
\end{equation}
where $\hat{P}^J_{M,K}$ is a projection operator~\cite{lap1,lap2,ring_book}. Note that while $K$ depends upon the orientation of the initial pair condensate, we can fix $M$ to be the same for all states.  Because these states are not necessarily 
orthonormal, we introduce a norm matrix,
\begin{equation}
{\cal N}^{J,M}_{\omega^\prime K^\prime, \omega K} = \langle \omega^\prime, J (K^\prime) M | \omega, J (K) M \rangle,
\end{equation}
as well as matrix elements of the Hamiltonian (here USDB~\cite{usdb}),
\begin{equation}
{\cal H}^{J,M}_{\omega^\prime K^\prime, \omega K} = \langle \omega^\prime, J (K^\prime) M | \hat{H}| \omega, J (K) M \rangle.
\end{equation}
We then solve the generalized eigenvalue problem,
\begin{equation}
\sum_{\omega, K} {\cal H}^{J,M}_{\omega^\prime K^\prime, \omega K} C^{J}_{\omega, K}(\epsilon) = E^J_\epsilon \sum_{\omega, K} {\cal N}^{J,M}_{\omega^\prime K^\prime, \omega K} C^{J}_{\omega, K}(\epsilon)
\end{equation}
where we label the eigenvalues for a given $J$ by $\epsilon$; because the Hamiltonian is an angular momentum scalar, the eigenvalues do not depend upon $M$. The eigenstates can be expanded,
\begin{equation}
| \epsilon,J \rangle = \sum_{\omega, K}  C^{J}_{\omega, K}(\epsilon) | \omega, J (K) M \rangle. 
\label{expansion}
\end{equation}

	\begin{table*}
		\caption{
			The projection of PVPC eigenstates with angular momentum $J$ within each subspace originating from minima with different ($\beta$, $\gamma$) deformations, 
			$\langle \epsilon; J|\beta,\gamma\rangle$, 
			 for $^{28}$Si, $^{26}$Si/Mg, $^{24}$Si/Ne, and $^{24}$Mg, which are highlighted in Fig. \ref{fig:land}(a) with shading. Based on the $\langle \epsilon; J|\beta,\gamma\rangle$ values, all eigenstates are categorized into specific bands with definite shapes or shape mixing. For triaxial deformations within $0<\gamma< 60^\circ$, a dual-band structure is anticipated, typically including two $2^+$, $4^+$, $6^+$ states, etc. The subscript of each eigenstate corresponds to its energy order in the PVPC levels with the same angular momentum. Because of the non-orthogonality of basis from different shapes,  $\sum_{(\beta,\gamma)}\langle \epsilon; J|\beta,\gamma\rangle^2\neq 1$.
		}\label{tab:pro}
		\begin{tabular}{cccc|cccccccccccccccccccc}
			\hline\hline									
			& $^{28}$Si & (0.195,60$^\circ$) & (0.229,0$^\circ$) &  & $^{26}$Si/Mg & (0.176,35$^\circ$) & (0.224,0$^\circ$) \\
			\hline
			\multirow{6}*{\rotatebox{90}{oblate}} & $0^+_1$ & 1.000 & $<$0.001 & \multirow{11}*{\rotatebox{90}{$\gamma\sim 35^{\circ}$}} & $0^+_1$ & 0.992 & 0.305 \\
			& $2^+_1$ & 1.000 & $<$0.001 &  & $2^+_1$ & 0.991 & 0.400 \\
			& $4^+_1$ & 0.999 & $<$0.001 &  & $4^+_1$ & 0.932 & 0.569 \\
			& $6^+_1$ & 0.859 & $<$0.001 &  & $6^+_2$ & 0.997 & 0.017 \\
			& $8^+_1$ & 0.998 & $<$0.001 &  & $8^+_2$ & 0.909 & 0.340 \\
			& $10^+_1$ & 0.998 & $<$0.001 &  &  &  &  \\
			&  &  &  &  & $2^+_2$ & 0.972 & 0.266 \\
			\multirow{6}*{\rotatebox{90}{prolate}} & $0^+_2$ & $<$0.001 & 1.000 &  & $4^+_3$ & 0.867 & 0.222 \\
			& $2^+_2$ & $<$0.001 & 0.997 &  & $4^+_3$ & 0.997 & 0.026 \\
			& $4^+_2$ & $<$0.001 & 0.997 &  & $6^+_3$ & 0.951 & 0.005 \\
			\cline{5-8}
			& $6^+_2$ & $<$0.001 & 0.996 & \multirow{5}*{\rotatebox{90}{prolate}} & $0^+_2$ & 0.067 & 0.952 \\
			& $8^+_2$ & $<$0.001 & 0.997 &  & $2^+_3$ & 0.271 & 0.877 \\
			& $10^+_2$ & $<$0.001 & 0.997 &  & $4^+_2$ & 0.619 & 0.791 \\
			&  &  &  &  & $6^+_1$ & 0.768 & 0.812 \\
			&  &  &  &  & $8^+_1$ & 0.522 & 0.940 \\
			\hline
			& $^{24}$Mg & (0.273,10$^\circ$) & (0.211,60$^\circ$) &  & $^{24}$Si/Ne & (0.162,32$^\circ$) & (0.171,60$^\circ$) \\
			\hline
			\multirow{8}*{\rotatebox{90}{$\gamma\sim 10^{\circ}$}} & $0^+_1$ & 1.000 & 0.576 & \multirow{8}*{\rotatebox{90}{$\gamma\sim 32^{\circ}$}} & $0^+_1$ & 0.987 & 0.995 \\
			& $2^+_1$ & 1.000 & 0.297 &  & $2^+_1$ & 0.997 & 0.889 \\
			& $4^+_1$ & 0.983 & 0.557 &  & $4^+_1$ & 0.993 & 0.826 \\
			& $6^+_1$ & 0.969 & 0.573 &  & $6^+_1$ & 0.998 & 0.880 \\
			&  &  &  &  &  &  &  \\
			& $2^+_2$ & 1.000 & 0.432 &  & $2^+_2$ & 0.996 & 0.443 \\
			& $4^+_2$ & 0.992 & 0.516 &  & $4^+_3$ & 0.977 & 0.433 \\
			& $6^+_2$ & 0.983 & 0.200 &  & $6^+_2$ & 0.889 & 0.605 \\
			\hline
			\multirow{4}*{\rotatebox{90}{oblate}} & $0^+_2$ & 0.005 & 0.818 & \multirow{3}*{\rotatebox{90}{mixing}} & $0^+_2$ & 0.162 & 0.104 \\
			& $2^+_3$ & 0.034 & 0.852 &  & $2^+_3$ & 0.126 & 0.114 \\
			& $4^+_3$ & 0.157 & 0.840 &  & $4^+_2$ & 0.878 & 0.520 \\
			& $6^+_3$ & 0.228 & 0.904 &  &  &  &  \\
			\hline\hline		
		\end{tabular}
	\end{table*}

To build an understanding of the eigenstates, we would like to trace back the contributions from the different minima, e.g., different $\omega = (\beta, \gamma)$. 
By orthonormalizing the set $\{ | \omega, J (K) M \rangle \}$, we can compute the projection of an eigenstates $| \epsilon, J \rangle$ on $ | \omega = (\beta, \gamma) \rangle$, that is, 
compute $ \langle \epsilon, J | (\beta, \gamma) \rangle$, the phase of which is unimportant. Note that because states of different shapes are not necessarily orthogonal, 
we can have $ \sum_{\beta,\gamma} |  \langle \epsilon, J | \omega=(\beta,\gamma) \rangle|^2 \neq 1$.  For more details see the Appendix.
	
	We list calculated $\langle \epsilon; J|\beta,\gamma\rangle$ for the eigenstates of $^{28}$Si, $^{26}$Si/Mg, $^{24}$Si/Ne, and $^{24}$Mg in Table \ref{tab:pro}, and assign each eigenstate to a certain band with a definite shape or shape-mixing nature. For instance, the $0^+_1$ state of $^{28}$Si exhibits a projection value of 1.000 in the subspace with a deformation of $(0.195,60^\circ)$, indicating that the $0^+_1$ state can be entirely expressed with the angular-projected basis from a nuclear intrinsic state with single oblate shape. 
	Conversely, the $4^+_2$ state of $^{24}$Si/Ne has projections of 0.878 and 0.520, respectively, in the subspaces originating from a triaxial state ($\gamma\sim 32^\circ$ ) and an oblate shape, with 
	strong shape-mixing. 

\begin{figure}
		\includegraphics[angle=0,width=0.45\textwidth]{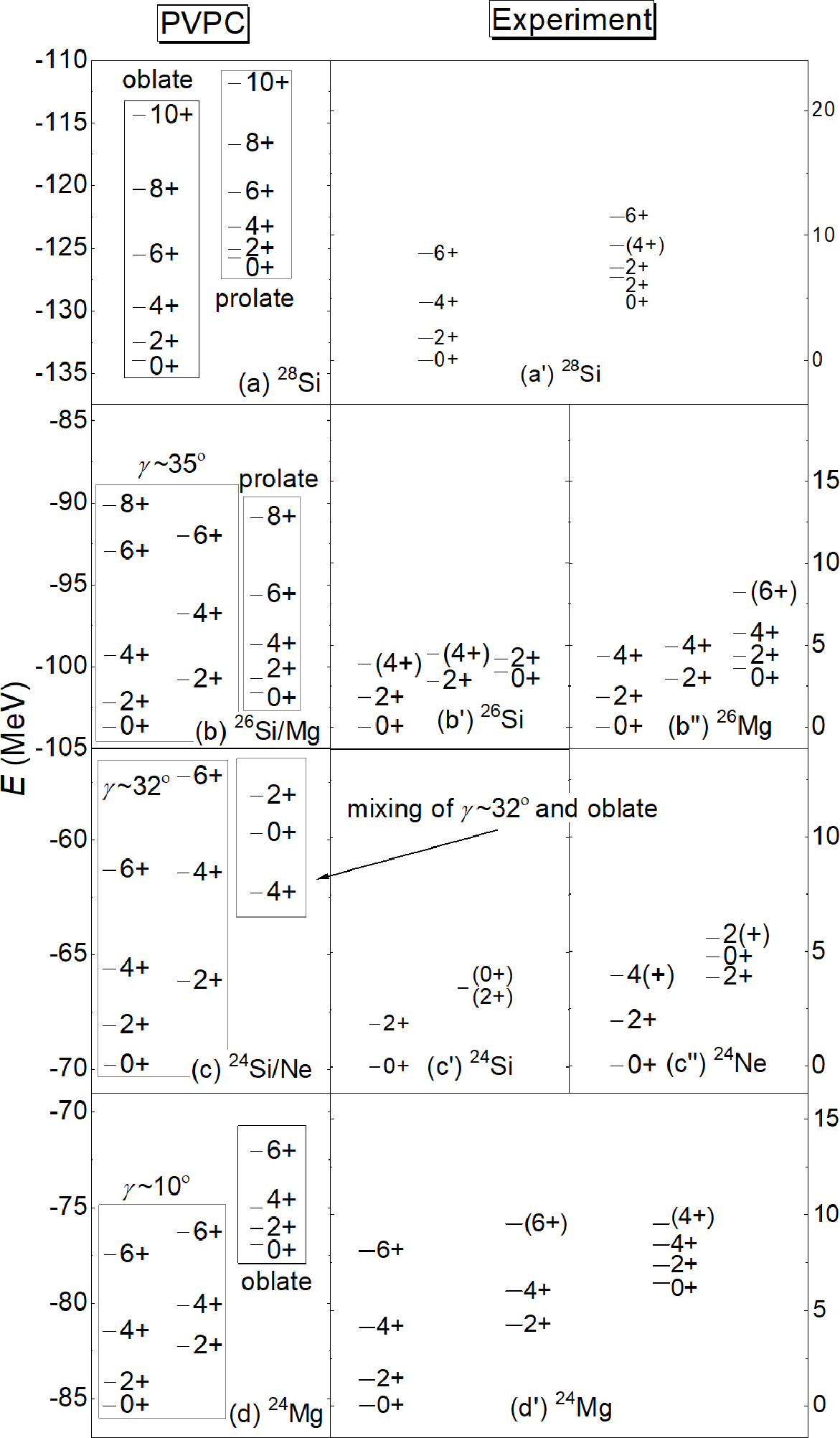}
		\caption{		
			Level schemes obtained from PVPC calculations alongside experimental data for $^{28}$Si \cite{28Si_level_exp}, $^{26}$Si \cite{26Si_level_exp}, $^{26}$Mg \cite{26Mg_level_exp}, $^{24}$Si \cite{24Si_level_exp}, $^{24}$Ne \cite{24Ne_level_exp}, and $^{24}$Mg \cite{24Mg_level_exp}, which are highlighted in Fig. \ref{fig:land}(a) with diagonal hatching. The shape assignments are detailed in Table \ref{tab:pro}, here highlighted with boxes.
		}\label{fig:spe}
	\end{figure}

	With the band assignments from Table \ref{tab:pro}, we present the calculated low-lying band structures of $^{28}$Si, $^{26}$Si/Mg, $^{24}$Si/Ne, and $^{24}$Mg obtained from PVPC, alongside experimental level schemes \cite{28Si_level_exp,26Si_level_exp,26Mg_level_exp,24Si_level_exp,24Ne_level_exp,24Mg_level_exp} in Fig. \ref{fig:spe}. In panels (a), (b), (c), and (d) , the assigned band structures are highlighted with boxes. Coexistence comes into play with bands originating from different minima, while	
$^{24}$Si/Ne contains a shape-mixing band.
	
	\begin{table*}
		\caption{
			Reduced E2 transition rates in Weisskopf units (W.u.) obtained from experiment and PVPC calculations for $^{28}$Si \cite{28Si_exp_nds}, $^{26}$Si/Mg \cite{26_exp_nds}, $^{24}$Si/Ne, and $^{24}$Mg \cite{24_exp_nds}, which are highlighted in Fig. \ref{fig:land}(a) with shading. The PVPC calculations adopt effective charges of $e_{\pi}=1.5e$ and $e_{\nu}=0.5e$. The subscript of each state corresponds to its energy order in PVPC results, as depicted in Fig. \ref{fig:spe}(a), (b), (c), and (d).
		}\label{tab:e2}
		\begin{tabular}{m{6em}<\centering m{6em}<\centering m{6em}<\centering |m{6em}<\centering m{6em}<\centering m{6em}<\centering m{6em}<\centering m{6em}<\centering  m{6em}<\centering}
			\hline\hline																															
			&	$^{28}$Si	&	PVPC	&										&	$^{26}$Si	&	PVPC	&	$^{26}$Mg	&	PVPC	\\
			\hline																															
			$	2	^+_	1	\rightarrow	0	^+_	1	$	&	13.2(5)	&	24.53	&	$	2	^+_	1	\rightarrow	0	^+_	1	$	&	15.3(15)	&	8.78	&	13.4(6)	&	14.38	\\
			$	4	^+_	1	\rightarrow	2	^+_	2	$	&	16.4(18)	&	33.11	&	$	4	^+_	1	\rightarrow	2	^+_	1	$	&		&	15.15	&	4.5(3)	&	16.15	\\
			$	6	^+_	1	\rightarrow	4	^+_	1	$	&	10.6(10)	&	32.64	&										&		&				&		\\
			$	8	^+_	1	\rightarrow	6	^+_	1	$	&		&	28.37	&	$	4	^+_	3	\rightarrow	2	^+_	2	$	&		&	0.64	&	2.5(6)	&	0.92	\\
			&										&		&				&		\\
			$	2	^+_	2	\rightarrow	0	^+_	2	$	&		&	25.70	&	$	2	^+_	3	\rightarrow	0	^+_	2	$	&		&	20.38	&		&	16.75	\\
			$	4	^+_	2	\rightarrow	2	^+_	3	$	&	11.1(18)	&	34.70	&										&		&				&		\\
			$	4	^+_	2	\rightarrow	2	^+_	2	$	&	30(4)	&	34.70	&	$	2	^+_	2	\rightarrow	0	^+_	1	$	&	1.7(4)	&	5.08	&	0.39(4)	&	0.19	\\
			$	6	^+_	2	\rightarrow	4	^+_	2	$	&	37(11)	&	34.20	&	$	2	^+_	2	\rightarrow	2	^+_	1	$	&	25(24)	&	12.07	&	6.1(21)	&	6.40	\\
			
			$	8	^+_	2	\rightarrow	6	^+_	2	$	&		&	29.71	&	$	4	^+_	3	\rightarrow	2	^+_	1	$	&	$>$2.3	&10.71		&	14(3)	&4.69		\\	
			&    &  &  &  &  &  &  \\
			$0^+_2\rightarrow 2^+_1 $ & 0.267(19)  & $<0.01$  & $0^+_2\rightarrow2^+_2$ & 10(5)  & 2.52 &  & 0.97 \\
			$2^+_2\rightarrow0^+_1$ & 0.37(15)  & $<0.01$  & $4^+_2\rightarrow2^+_2$ &  & 0.31 & 1.7(5) & 0.06 \\
			$2^+_3\rightarrow0^+_1$ & 0.162(18)  & $<0.01$  & $2^+_3\rightarrow0^+_1$ & 0.28(15) & 0.24 &  & $<0.01$ \\
			$4^+_2\rightarrow2^+_1$ & 0.084(11)  & $<0.01$  & $0^+_2\rightarrow2^+_1$ &  & 3.01 & 1.07(8) & 0.64 \\
			$6^+_2\rightarrow4^+_1$ & 0.17(5)  & $<0.01$  & $2^+_3\rightarrow0^+_1$ &  & 0.24 & 0.24(5) & $<0.01$ \\
			\hline																															
			&	$^{24}$Mg	&	PVPC	&										&	$^{24}$Si	&	PVPC	&	$^{24}$Ne	&	PVPC	\\
			\hline																															
			$	2	^+_	1	\rightarrow	0	^+_	1	$		&		$21.07^{+48}_{-46}$		&		17.60		&		$	2	^+_	1	\rightarrow	0	^+_	1	$		&		4.6(14)		&		9.72		&				&		8.67		\\
			$	4	^+_	1	\rightarrow	2	^+_	1	$		&		$35.7^{+34}_{-29}$		&		22.03		&		$	4	^+_	1	\rightarrow	2	^+_	1	$		&				&		10.97		&				&		10.58		\\
			$	6	^+_	1	\rightarrow	4	^+_	1	$		&		$38^{+18}_{-10}$		&		19.43		&		$	6	^+_	1	\rightarrow	4	^+_	1	$		&				&		13.19		&				&		9.23		\\
			&				&				&		&		&		&		&		\\
			$	4	^+_	2	\rightarrow	2	^+_	2	$		&		14.9(12)		&		9.31		&		$	4	^+_	3	\rightarrow	2	^+_	2	$		&				&		7.70		&				&		5.37		\\
			$6^+_2\rightarrow 4^+_2$   & $24^{+22}_{-9}$   & 10.49 & $6^+_2\rightarrow 4^+_3$ &   & 9.48 &  & 4.98 \\
			&  &  &  &  &  &  &  \\
			$2^+_3\rightarrow0^+_2$ &  & 11.35 & $2^+_2\rightarrow0^+_1$ &  & 0.56 &  & 0.28 \\
			$4^+_3\rightarrow2^+_3$ & $59^{+43}_{-19}$ & 7.81 & $2^+_2\rightarrow2^+_1$ &  & 15.18 &  & 12.36 \\
			&  &  & $2^+_2\rightarrow4^+_1$ &  & 0.6 &  & 1.23 \\
			$2^+_2\rightarrow0^+_1$ & $1.72^{+14}_{-12}$ & 1.1 &  &  &  &  &  \\
			$2^+_2\rightarrow2^+_1$ & 3.36(27) & 2.3 & $4^+_3\rightarrow2^+_1$ &  & 3.02 &  & 0.3 \\
			$4^+_2\rightarrow2^+_1$ & 1.06(8) & 1.24 & $4^+_3\rightarrow4^+_1$ &  & 6.7 &  & 3.41 \\
			$6^+_2\rightarrow4^+_1$ & $0.8^{+7}_{-3}$ & 1.37 & $4^+_3\rightarrow6^+_1$ &  & 0.89 &  & 0.77 \\
			&  &  &  &  &  &  &  \\
			$0^+_2\rightarrow2^+_2$ & $6.8^{+15}_{-11}$ & 1.35 & $6^+_2\rightarrow4^+_1$ &  & 0.02 &  & 0.2 \\
			$4^+_3\rightarrow2^+_2$ & $2.1^{+15}_{-6}$ & 1.3 & $6^+_2\rightarrow6^+_1$ &  & 4.74 &  & 2.99 \\
			$0^+_2\rightarrow2^+_1$ & $0.50^{+10}_{-7}$ & 0.35 &  &  &  &  &  \\
			$2^+_3\rightarrow0^+_1$ & $0.61^{+31}_{-15}$ & 0.01 & $0^+_2\rightarrow2^+_1$ &  & 0.12 & 0.46(16) & 0.33 \\
			$4^+_3\rightarrow2^+_1$ & $1.6^{+12}_{-5}$ & 0.03 & $0^+_2\rightarrow2^+_2$ &  & 17.91 &  & 13.16 \\
			\hline\hline
		\end{tabular}
	\end{table*}
	
	In addition to excitation spectra, in Table~\ref{tab:e2} we  present  the reduced E2 transition rates, denoted as B(E2), obtained from experiments \cite{28Si_exp_nds,26_exp_nds,24_exp_nds} and from our PVPC calculations; the latter adopt  effective charges of $e_{\pi}=1.5e$ and $e_{\nu}=0.5e$. We include intraband transitions as well as some noteworthy interband transitions, using
	 band assignments from Table \ref{tab:pro}.
	
	The well-established prolate-oblate coexistence in $^{28}$Si shows up as two rotational bands in Fig.~\ref{fig:spe}(a) and (a$^\prime$). The calculated moments of inertia of these two bands are consistent with experiment. The calculated intraband B(E2) values in Table \ref{tab:e2} are approximately 30 W.u., consistently larger than the experimental B(E2) of the  oblate ground state band, yet in agreement with those of the prolate band, suggesting that our calculation overestimates the collectivity of the ground band. The calculated interband B(E2) values are less than 0.001 W.u., significantly smaller  than the experimental values (on the order of $\sim$0.1 W.u.), demonstrating  weaker mixing of the two shapes in $^{28}$Si than is observed experimentally. The lack of mixing is supported by 
Table \ref{tab:pro}, where 	eigenstates in the oblate band have projections of less than 0.001 in the prolate subspace, and vice versa.  Therefore, while the intraband transitions and level scheme of the prolate-oblate coexistence appear reasonable, our PVPC calculation nonetheless underestimates the prolate-oblate mixing in $^{28}$Si. We note that the the coexistence of oblate and prolate shapes in $^{28}$Si using the USDB interaction was also recently reported by Ref. \cite{rec_prc}, which further corroborates our observations regarding shape coexistence in $^{28}$Si.
	
	In Fig.~\ref{fig:spe}(b), (b$^\prime$), and (b$^{\prime \prime}$), we present the low-lying level schemes of $^{26}$Si/Mg. In the PVPC results, a triaxial-oblate coexistence is evident, corresponding to the ground band and its side band with $\gamma=35^\circ$, alongside an prolate rotational band starting from the excited $0^+$ state at $\sim 2.5$MeV, as illustrated in Fig.~\ref{fig:spe}(b). The E2 transition scheme summarized in Refs. \cite{26_exp_nds} also reveals three band structures in both $^{26}$Si and $^{26}$Mg, as shown in Fig.~\ref{fig:spe}(b$^\prime$) and (b$^{\prime \prime}$), consistent with the picture of triaxial-prolate coexistence. While there is reasonable agreement between theoretical and experimental band structures, the moment of inertia of the triaxial side band is slightly underestimated in the PVPC results. Notably, the strong collectivity of the ground band is suggested by experimental B(E2, $2^+_1\rightarrow 0^+_1$) values of $\sim$15 W.u. for both $^{26}$Si and $^{26}$Mg, which our calculations reproduce within a factor of 2. Strong interband transitions between the triaxial ground band and the side band, such as $2^+_2\rightarrow 2^+_1$ and $4^+_3\rightarrow 2^+_1$, indicating $K$-mixing between triaxial bands, are observed in the experimental data and reproduced by our calculation. However, a notable discrepancy arises in the transition from the prolate $0^+_2$ state to the triaxial $2^+_2$ state, where the experimental B(E2) value is 10(5) W.u., while  PVPC yields 2.52 W.u.. In Table~\ref{tab:pro} the $0^+_2$ state is primarily a prolate configuration, while the $2^+_2$ state is predominantly triaxial. Although the $2^+_2$ state also contains a nontrivial component (0.266 projection) in the prolate subspace, due to the non-orthogonality between triaxial and prolate subspaces, this is  insufficient to reproduce the strong transition rate of $0^+_2\rightarrow 2^+_2$. Therefore, similarly to  $^{28}$Si, it appears that  PVPC underestimates the shape mixing in $^{26}$Si, and possibly in $^{26}$Mg as well.
	
	In the computed level scheme of $^{24}$Si/Ne in Fig.~\ref{fig:spe}(c), three band structures emerge. The ground band and a side band are from a triaxial state with $\gamma\sim 32^\circ$, while the third band arises from a mixing of the triaxial and oblate minima as detailed in Table~\ref{tab:min}. Table~\ref{tab:pro} also reveals that yrast band members also exhibit a $\sim 0.8$ projection in the oblate subspace, despite having projections in the triaxial subspace exceeding 0.98, indicating a significant non-orthogonality between the triaxial and oblate minima in $^{24}$Si/Ne. Consequently, the mixing of the third band results in an irregular level scheme with a $0^+$ state situated between $2^+$ and $4^+$ states, and a $2^+$ state higher in energy than the $4^+$ state. However, the moment of inertia of the ground band is well reproduced by the PVPC. Notably, the experimental  B(E2,$2^+_1\rightarrow 0^+_1$) value  is 4.6(14) W.u., a value not far from  the PVPC result of 9.72 W.u., as found in Table~\ref{tab:e2}. It is important to mention that the experimental ``yrare" level schemes of $^{24}$Si and $^{24}$N appear to be incomplete, with uncertain spin-parity assignments, as depicted in Fig.~\ref{fig:spe}(c$^\prime$) and (c$^{\prime \prime}$). Consequently, it is challenging to evaluate the agreement of the triaxial side band or the mixing band with the experimental ``yrare" band. Surprisingly, although the calculated $0^+_2$ level resides in an irregular energy region, the PVPC yields an interband B(E2,$0^+_2\rightarrow 2^+_1$) close to the experimental value. The strong interband transitions between the ground band and the ``yrare" band, such as $2^+_2\rightarrow 2^+_1$, $4^+_3\rightarrow 4^+_1$, $6^+_2\rightarrow 6^+_1$, and $0^+_2\rightarrow 2^+_2$, support a strong mixing for the ``yrare" band.
	
	Lastly, in $^{24}$Mg, we identify two distinct VPC minima, a near-prolate deformation with $\gamma\sim 10^\circ$ and a oblate deformation, as listed in Table \ref{tab:min}, leading to  three distinct band structures; see Table~\ref{tab:pro} and Fig.~\ref{fig:spe}(d). The ground band and a side band arise from the slightly triaxial state: the PVPC results have  moments of inertia consistent with experiment \cite{24Mg_exp}, with calculated intraband and interband B(E2) values that agree well with experiment, as detailed in Table \ref{tab:e2}.  In addition, we find a  third,oblate band, corresponding to the experimental $0^+_2$ state at 6.432 MeV, as expected from Table III of Ref. \cite{coexistence_rev_the_rmp}. The experimental E2 transition strengths for this oblate band, including $4^+_3\rightarrow 2^+_2$, $0^+_2\rightarrow 2^+_1$ and so on, are well reproduced by  PVPC values in Table \ref{tab:e2}. Note that despite considerable efforts to investigate the $0^+_2$ state \cite{24Mg_exp_0_plb,24Mg_exp_0_pro,24Mg_exp}, its collective nature remains inconclusive due to the lack of experimentally identified higher spin states in this oblate band. The PVPC calculation depicted in Fig. \ref{fig:spe}(d) also does not yield $8^+$ or higher spin eigen-states purely from the oblate minimum, suggesting that this oblate band may terminate at the $6^+$ state.
	
	\section{summary}\label{sec:sum}
	
	In summary, our investigation employs the VPC method with generalized pairing to explore  shape coexistence across several nuclear regions,  up to  50-82. As depicted in Fig.~\ref{fig:land}, the majority of nuclei exhibit two or more minima, underscoring the pervasiveness of coexistence throughout the nuclide chart. In general, heavy nuclei or those near the drip line tend to feature a larger number of coexisting minima.
	
	Coexistence is sensitive to degrees of freedom:  TRVPC calculations, imposing time-reversal constraints on pairing, and HF calculations, which exclude all pairing correlation, yield fewer minima than our most general PVPC calculations. In particular, this underscores the importance of allowing all possible pairing channels in VPC to capture shape coexistence.
	
	To express the role of  generalized pairing, we introduced a one-body entropy which measures the deviation of our condensates from a single Slater determinant. We found that  energy is more sensitive to entropy that deformation, and the lowest minimum is likely to have higher entropy (i.e., more pairing correlation) than higher local minima. We also found that large entropy emerges near shell closure, where one might expect pairing to dominate over deformation.
	
	To illustrate the impact of potential coexistence on low-lying structure, we projected out states of good angular momentum and studied in detail the bands found in $^{28}$Si, $^{26}$Si/Mg, $^{24}$Si/Ne, and $^{24}$Mg.  We reproduce the known coexistence in $^{28}$Si, aligning well with experimental data and some prior theoretical studies. Additionally, we propose new instances of coexistence and shape mixing in $^{26}$Si/Mg and $^{24}$Si/Ne, and suggest their experimental correspondence with level-scheme and transition-rate observations. Confirming the potential coexistence of $^{24}$Mg of near-prolate and oblate deformation, our analysis indicates that the oblate band starting from the $0^+$ state at 6.432 MeV may potentially culminate in a $6^+$ state. Through scrutinizing low-lying wave-functions, level schemes, and transition patterns, our investigation underscores the reliability of the coexistence picture derived from our VPC calculations.

	\begin{acknowledgments}
		This work has received support from various sources, including the National Natural Science Foundation of China (Grants No. 12105234, No. 11705100, No. 12175115, and No. 11875188), the Natural Science Foundation of Shandong Province, China (ZR2022MA050), and the U.S. Department of Energy, Office of Science, Office of Nuclear Physics, under Award No. DE-FG02-03ER41272.
	\end{acknowledgments}
	
	\appendix
	
	\section{Projecting onto shape subspaces}
	
	\label{app:project}
	
	For a physical interpretation of the band structure of a given nucleus, we would like to compute the projection of each state into a subspace defined by a minimum with the deformation parameters $(\beta,\gamma)$.
Because the basis states in Eq.~(\ref{expansion}) are not necessarily orthonormal,  this task is non-trivial. In this appendix we discuss both the non-orthogonality within a subspace spanned by the 
projected states for a given shape $(\beta,\gamma) \equiv \omega$, as well as the non-orthogonality between different shape subspaces.

We first consider the projected states $| \omega, J(K) M \rangle $ defined in Eq.~(\ref{projection}). To define a shape subspace, we fix $\omega $.  We also fix total angular momentum $J$ and laboratory frame $z$-component of angular momentum $M$, as states which differ in these quantum numbers must be orthogonal. Thus we introduce a similarity transformation $G^{\omega, J}_{g,K}$ 
which diagonalizes the subblock of the norm matrix $ \langle \omega, J(K) M | \omega, J(K^\prime) M \rangle $, that is, let 
\begin{equation}
| \omega, g, JM \rangle = \sum_K G^{\omega, J}_{g,K}  | \omega, J(K) M \rangle
\end{equation}
such that 
\begin{equation}
\langle \omega,g, JM | \omega, g^\prime, JM \rangle = \delta_{g,g^\prime}.
\end{equation}
Note that, since $M$ is just the laboratory frame orientation, the transformation is independent of $M$ as long as $|M| \leq J$. 

Now we can transform Eq.~(\ref{expansion}) into this new basis:
\begin{equation}
| \epsilon,J \rangle = \sum_{\omega, K}  \tilde{C}^{J}_{\omega, g}(\epsilon) | \omega, g, J M \rangle,
\label{expansion2}
\end{equation}
where 
\begin{equation}
 \tilde{C}^{J}_{\omega, g}(\epsilon) = \sum_K  G^{\omega, J}_{g,K} {C}^{J}_{\omega, K}(\epsilon).
\end{equation}
Finally, we can unambiguously define the project of the wave function into a given shape subspace by 
\begin{equation}
\langle \epsilon, J | \omega  \rangle = \sqrt{ \sum_g  \left |    \tilde{C}^{J}_{\omega, g}(\epsilon)   \right |^2   } .
\end{equation}
So, for example, if a state labeled by $\epsilon$ is entirely constructed from a given VPC minimum with deformation parameters $(\beta,\gamma) = \omega$, then 
 $\langle \epsilon, J | \omega  \rangle=1$.

We still have to consider that the deformation parameters $\beta,\gamma$ are not quantum numbers, that is, they are not eigenvalues, but are computed from diagonalizing the expectation value of the quadrupole tensor. Thus minima with different deformation parameters are not necessarily orthogonal, and hence one can have
\begin{equation}
\sum_\omega | \langle \epsilon, J | \omega  \rangle |^2 \neq 1. 
\end{equation}

	%
	
\end{document}